\documentclass[12pt,a4paper,english,superscriptaddress,aps,nofootinbib]{revtex4}
\usepackage{amsmath,amssymb,graphicx}
\makeatletter
\usepackage{babel}
\usepackage[active]{srcltx}
\usepackage{graphicx,color}
\usepackage{changebar}
\usepackage{hyperref}
\hypersetup{
	colorlinks=true,
	urlcolor= blue,
	citecolor=blue,
	linkcolor= blue,
	bookmarks=true,
	bookmarksopen=false,}

\usepackage[T1]{fontenc}
\usepackage{ae}
\usepackage[ansinew]{inputenc}
\usepackage{amsmath}
\usepackage{braket}
\usepackage{mathtools}
\usepackage{slashed}
\usepackage{empheq}
\usepackage{tikz}
\usepackage{multirow}
\usetikzlibrary{decorations.pathmorphing}
\usepackage[caption = false]{subfig}

\usepackage{graphicx}
\usepackage{amsfonts}
\usepackage{amsmath}
\begin{document}

\title{\Large{\bf Possible Lorentz symmetry violation from broken Weyl invariance} }

\author{Feng Wu}
\email[]{E-mail: fengwu@ncu.edu.cn}
\affiliation{%
Department of Physics, Nanchang University,
330031, China}

\begin{abstract}
\vspace{0.25cm}
\begin{center}
    \large{Abstract}
\end{center}
\vspace{0.01cm}
In this work, we investigate a theory of linear Weyl gravity coupled to a scalar field and study the scenario in which Lorentz symmetry is broken by a non-vanishing vacuum expectation value of the Weyl field in the flat space limit after Weyl symmetry breaking. We show that a $CPT$-odd Lorentz-violating interaction is generated after symmetry breaking. Features of different symmetry-broken phases and their dependence on the spacetime character of the generated Lorentz-violating background are discussed. Also, we analyze the naturalness of the theory by showing that the light mass scale is protected from large radiative corrections due to an enhanced spacetime symmetry.
\end{abstract}

\maketitle
\thispagestyle{empty}
\newpage

\section{Introduction}
When building an action to describe the dynamics of a physical system, symmetries underlying it play a vital role since they put restrictions on the form of the action. On the other hand, when attempting to extend the existing physical model, exploring the possibility of the absence of some symmetries is one of the effective approach.

Our present understanding of particle physics is well-described by the standard model (SM) of particle physics. However, as an effective field theory, the SM is incomplete and leaves a number of puzzles unresolved. One of them, for example, is why there are widely separated mass scales in nature. Because of this hierarchy problem, it is reasonable to expect that the signature of new physics should emerge at scales not far beyond the TeV scale. Thus, exploring the extension of the SM in particle physics has become an extensively investigated subject. In particular, as a potential new physics, the breaking of Lorentz symmetry and its related contents have been intensively studied since theoretically there is no good reason to believe that Lorentz invariance would be intact at all energies.\footnote{See, for instance, Refs. \cite{LV1,LV2,LV3,LV4,LV5,LV6,LV7} for works on the SM modified by Lorentz-violating interactions.}

In this paper, we will explore the effects of the Lorentz violation emerging from symmetry breaking in Weyl geometry. Weyl geometry originated from Hermann Weyl's attempt to geometrize the electromagnetic interaction \cite{Weyl}. In Weyl geometry, the constraint of length conservation is relaxed and there is no absolute scaling for spacetime, so that the concept of length is path-dependent. Therefore, different from the situation in Riemanian geometry, parallel transport of two vectors along a spacetime curve in Weyl geometry will change both their lengths and orientations. Only relative lengths and relative orientations remain fixed. The vector potential $A_{\mu}(x)$ in electrodynamics was identified with the gauge field in this geometry. However, Weyl's original idea was unsuccessful since it leads to conflict with physical facts, as pointed out by Einstein. For example, the mass of a free particle would depend on its history in Weyl's approach.

Weyl symmetry must be broken in the infrared. Recently, it has been shown that the Hilbert-Einstein action can be obtained from the matterless part of Weyl gravity after the symmetry breakdown of Weyl symmetry via either Stueckelberg mechanism \cite{Gh1, Gh2} or Coleman-Weinberg mechanism \cite{Oda1,Oda2}. In other words, the Hilbert-Einstein action could be a symmetry-broken phase of the Weyl action and the Planck scale is an emerging scale generated via these mechanisms.\footnote{The Planck scale generated from models governed by a global Weyl symmetry instead of a local one was discussed in \cite{Ferreira1, Hill1, Ferreira2}.} Note that Coleman-Weinberg mechanism arises from the explicit breaking of scale symmetry via the trace anomaly, which is turned on by quantum loops and is proportional to the $\beta$-function of the running coupling constant. Thus, quantum mechanics furnishes the trace anomaly and then induces the generation of mass scales. This mass generating mechanism is a pure quantum effect.\footnote{See Ref. \cite{Hill} for the elaboration of this point and the role of the renormalization group in this mechanism.} The above-mentioned works focus on the symmetry breakdown of Weyl symmetry and its related issues. In this work we consider the scenario in which the Weyl gauge field is the ``order parameter" such that the Weyl symmetry breaking is triggered by a non-vanishing fixed vacuum expectation value $\omega_{\mu}$ of the Weyl gauge field at some high energy scale. This fixed vector background then becomes the source that induces the Lorentz violation.\footnote{For some aspects of Lorentz violation due to aetherlike fields, see Refs. \cite{Gomes,Lan}.} The fact that a non-vanishing vacuum expectation value of the Weyl gauge field could be realized in the Weyl broken phase has been shown in \cite{Oda2}. However, the physical consequences derived from this Lorentz-violating background at low energies have not been explored. In this paper, we take steps towards studying some of them. We restrict ourselves to investigate a specific model of linear Weyl gravity coupled to a scalar field. This model has the virtue of simplifying the analysis without discarding the essential features we would like to address.

The rest of this paper is organized into four parts. In the next section, after describing the model we want to explore, several distinct features of the model are addressed. Then we go on to show that a $CPT$-odd Lorentz-violating interaction appears after Weyl symmetry breaking triggered by a non-vanishing $\omega_{\mu}$. Theoretical consequences are studied. In particular, we investigate different phases after Weyl symmetry breaking, depending on the spacetime character of the fixed background $\omega_{\mu}$. We show that the breakdown of Weyl symmetry does not necessarily generate a mass for the Weyl gauge field. After these analyses, the stability of the light mass scale generated after spontaneous symmetry breaking is clarified in Sect. 3. In Sect. 4, based on the analysis of Sect. 2, we consider the quantum corrections and calculate the one-loop effective potentials, for $\omega_{\mu}$ to be timelike and lightlike, respectively, and discuss their physical consequences. Our conclusions are presented in the final section.

\section{Lorentz violation after Weyl symmetry breaking}
Our starting point is the following action of linear Weyl gravity coupled to a scalar field $\phi(x)$:
\begin{equation}
S=\int d^4 x \sqrt{- g} \{ {\alpha\over 12}\phi^2 \tilde{R} + {1\over 2} g^{\mu\nu} D_{\mu} \phi D_{\nu} \phi - {1\over 4!}\lambda \phi^4-{1\over 4} F_{\mu\nu}^{2} \}, \,\,\,\,\,\,\,\,\,\,\,\, \lambda >0 .\label{Action}
\end{equation}
Here $D_{\mu} \phi  $ is the Weyl-covariant derivative of $\phi$ whose explicit form is defined as
\begin{equation}
D_{\mu} \phi \equiv (\partial_{\mu}-f S_{\mu})\phi
\end{equation}
with $S_{\mu}(x)$ the Weyl gauge field and $f$ the coupling constant to $\phi(x)$. $F_{\mu\nu}\equiv \partial_{\mu}S_{\nu}-\partial_{\nu}S_{\mu} $ is the field tensor for the Weyl field $S_{\mu}$. The Weyl scalar curvature $\tilde{R} $ is related to the Riemannian scalar curvature $R$ by
\begin{equation}
\tilde{R}=R -6f \nabla_{\mu} S^{\mu} -6 f^2 S_{\mu} S^{\mu}. \label{tildeR}
\end{equation}

The action~(\ref{Action}) is invariant under the Weyl transformations:
\begin{eqnarray}
g_{\mu\nu}(x) &\rightarrow& g_{\mu\nu}'(x)=e^{-2\Lambda (x)} g_{\mu\nu}(x), \,\,\,\,\,\,\,\, \phi(x) \rightarrow \phi'(x)=e^{+\Lambda (x)}  \phi(x),\nonumber \\
S_{\mu}(x) &\rightarrow& S_{\mu}'(x) = S_{\mu}(x)+{1\over f} \partial_{\mu} \Lambda (x). \label{trx}
\end{eqnarray}
In this paper we work with the metric signature $(+,-,-,-)$. We remark that the transformation of the Weyl gauge field $S_{\mu}$ in~(\ref{trx}) is different from that of the electromagnetic vector potential under $U(1)$ gauge symmetry due to the nonoccurrence of the imaginary unit $i$ in the Weyl covariant derivative. Thus, while the $U(1)$ gauge group in electrodynamics is compact, the symmetry group generated by the Weyl transformations~(\ref{trx}) is non-compact. When the Weyl gauge field takes the pure gauge form, the Weyl geometry is gauge-equivalent to the Riemann geometry. Also, we note in passing that it has been shown that with the Weyl symmetry, the Weyl field $S_{\mu}$ does not couple to the gauge and fermion fields of the SM \cite{Cesare, Gh3}.

Several comments are in order regarding the Weyl invariant action~(\ref{Action}) at the classical level. First, in Riemann geometry (i.e., setting the Weyl gauge field $S_{\mu}(x)=0$) the action~(\ref{Action}) reduces to
\begin{equation}
S_R=\int d^4 x \sqrt{- g} \{ {\alpha\over 12}\phi^2 R + {1\over 2} g^{\mu\nu} \partial_{\mu} \phi \partial_{\nu} \phi - {1\over 4!}\lambda \phi^4 \}.  \label{SR}
\end{equation}
In general, $S_R$ is not Weyl invariant. Indeed, it is easy to check that only when $\alpha=1$ will the action $S_R$ be Weyl invariant. In the case where $ \alpha=1$, the action $S_R$ is the well-known conformal scalar action.

Second, each term in~(\ref{Action}) is separately Weyl invariant. With $\alpha=1$, the $S_{\mu}$-dependent terms from $D_{\mu}\phi D^{\mu} \phi$ and from $\phi^2 \tilde{R}$ cancel each other, yielding a Weyl invariant action:
 \begin{equation}
S_C=\int d^4 x \sqrt{- g} \{ {1\over 12}\phi^2 R + {1\over 2} g^{\mu\nu} \partial_{\mu} \phi \partial_{\nu} \phi - {1\over 4!}\lambda \phi^4-{1\over 4} F_{\mu\nu}^{2} \}.  \label{SC}
\end{equation}
It is straightforward to show that the energy-momentum tensor $T_{\mu\nu}={-2\over \sqrt{-g}}{\delta S\over \delta_{g_{\mu\nu}}}\vert_{g_{\mu\nu}=\eta_{\mu\nu}}$ of the theory defined by $S_C$ takes the form
\begin{eqnarray}
T_{\mu\nu} = &-&{2\over 3} \partial_{\mu} \phi \partial_{\nu} \phi +{1\over 6} \eta_{\mu\nu} \partial_{\sigma} \phi \partial^{\sigma} \phi+{1\over 3}\phi (\partial_{\mu}\partial_{\nu}-\eta_{\mu\nu}\square)\phi -\eta_{\mu\nu} {1\over 4!}\lambda \phi^4 \\ \nonumber
&+&F_{\mu}^{\,\,\,\lambda}F_{\nu\lambda}-{1\over 4}\eta_{\mu\nu} (F_{\rho\sigma})^{2}.
\end{eqnarray}
The trace of $T_{\mu\nu} $ vanishes ``on-shell" (i.e., assuming the fields satisfy their equations of motion), as it should be for a conformal field theory \cite{Callan}.

Third, notice that in the action $S_{C}$, the Weyl field $S_{\mu}$ couples to the scalar field $\phi$ only through gravity, and thus the two fields must decouple in the flat limit. Explicitly, in the flat limit, $S_C$ reduces to two independent sectors:
\begin{equation}
S_C\longrightarrow S_\phi+ S_A =\int d^4 x \left( {1\over 2}  \partial_{\mu} \phi \partial^{\mu} \phi - {1\over 4!}\lambda \phi^4\right) + \int d^4 x' \left(-{1\over 4} F_{\mu\nu}^{2}\right) . \label{SF}
\end{equation}
This flat-limit action enjoys an extended Poincar$\rm\acute{e}$ symmetry \cite{Foot}, meaning that the scalar sector $S_\phi$ and the vector sector $S_A$ in Eq.~(\ref{SF}) are individually invariant under independent Poincar$\rm\acute{e}$ transformations. Therefore, for the coupling constant $\alpha$ in Eq.~(\ref{Action}), small values of $(\alpha-1)$ are protected by this emerging symmetry and are technically nature \cite{Hooft}, as long as we are in the flat limit so that the effects due to gravity that mixes the two sectors can be neglected. In the next section, we will show that after spontaneous symmetry breaking, it is this enhanced spacetime symmetry in the decoupling limit $\alpha\rightarrow 1$ that secures the stability of the light mass scale in the symmetry-broken Lagrangian. From now on, we will always assume that $(\alpha-1)$ is very small.

We now go back to the action~(\ref{Action}). It was shown in \cite{Oda2} that upon Weyl symmetry breaking, the vacuum expectation value of the Weyl scalar curvature $\langle \tilde{R}\rangle$ is non-vanishing. In order to have a flat metric ($R=0$) at low energies, Eq.~(\ref{tildeR}) guides us to assume that $S_{\mu}$ acquires a non-vanishing fixed vacuum expectation value $\omega_{\mu}$. This fixed background field, acting like an aether field, breaks the Lorentz invariance while leaves the translational invariance intact. For timelike, spacelike, and lightlike $\omega_{\mu}$, the lorentz group $SO(1,3)$ is broken down to the little groups $SO(3)$, $SO(1,2)$, and $E(2)$, respectively.

Let us define the shifted field by writing
\begin{equation}
S_{\mu}(x)=\omega_{\mu}+A_{\mu}(x)
\end{equation}
and expand the Lagrangian about $\omega_{\mu}$. In the flat limit, meaning that we neglect the gravitational effects, the Lagrangian density of Eq.~(\ref{Action}) reduces to
\begin{eqnarray}
\mathcal{L}&=&{1\over 2}\partial_{\mu} \phi(x) \partial^{\mu} \phi(x) -{1\over 2} m_{\phi}^{2} \phi^2(x) - {1\over 4!}\lambda \phi^4(x)  \nonumber\\
&&-{1\over 2} f (\alpha-1)\phi^2(x)(\partial_{\mu}+2f \omega_{\mu} )A^{\mu}(x)-{1\over 2} f^2 (\alpha-1)\phi^2(x) A_{\mu}^2(x)-{1\over 4} F_{\mu\nu}^{2}(x), \label{Lag}
\end{eqnarray}
where the mass squared parameter $ m_{\phi}^{2}$ of $\phi$ is defined as
\begin{equation}
m_{\phi}^{2}\equiv f^2(\alpha-1)\omega_{\mu}^2.
\end{equation}

Notice the appearance of the Lorentz-violating operator $\omega_{\mu} A^{\mu}(x) \phi^2(x)$ in Eq.~(\ref{Lag}). While this operator preserves $PT$, it violates $C$ parity. Thus, it is a $CPT$-odd operator. We note that $CPT$ invariance is a necessary but not sufficient condition for Lorentz invariance of an interacting quantum field theory.

The potential for $\phi$ in Eq.~(\ref{Lag}) takes the familiar form
\begin{equation}
V(\phi)={1\over 2} m_{\phi}^{2} \phi^2+ {1\over 4!}\lambda \phi^4.
\end{equation}
Whether the $Z_2$ symmetry ($\phi \rightarrow -\phi$) of the Lagrangian is preserved or spontaneously broken is determined by the location of the minimum of the potential. It is instructive to consider the three cases $m_{\phi}^{2}>0$, $m_{\phi}^{2}=0$, and $m_{\phi}^{2}<0$.\\

(i) $m_{\phi}^{2}>0$ (either $\alpha >1$ and $\omega_{\mu}$ timelike, or $\alpha <1$ and $\omega_{\mu}$ spacelike):\\

In this case, the potential $ V(\phi)$ has a minimum at $\phi=0$. Thus, after Weyl symmetry breaking, the scalar field $\phi$ acquires a mass while the ``Weyl photon" $A_{\mu}$ remains massless\footnote{However, when considering the quadratic Weyl gravity for which the term $\tilde{R}^2$ is included, the Weyl photon would acquire a mass squared $m_{A}^2 \sim f^4 \omega_{\mu}^2$, arising from the term $\tilde{R}^2$ after Weyl symmetry breaking. Then, the Weyl photon is massive for timelike $\omega_{\mu}$.}.\\

(ii) $m_{\phi}^{2}=0$ (i.e., $\omega_{\mu}$ lightlike):\\

In this case, even with non-vanishing $\omega_{\mu}$ after Weyl symmetry breaking, it appears that both fields remains massless. However, the mass of these fields can be generated dynamically from radiative corrections via the Coleman-Weinberg mechanism \cite{Coleman}. In Sect. 4, we will show that the Weyl gauge field $A_{\mu}$ can acquire a positive mass squared through this mechanism only if $\alpha<1$.\\

(iii) $m_{\phi}^{2}<0$ (either $\alpha <1$ and $\omega_{\mu}$ timelike, or $\alpha >1$ and $\omega_{\mu}$ spacelike):\\

In this case, the potential $V(\phi)$ has its minimum at $\phi_{c}=\pm ({-6m_{\phi}^2 \over \lambda})^{1/2}$. Thus, the $Z_2$ symmetry would further be spontaneously broken after Weyl symmetry breaking. Expanded around the broken-symmetry state by writing
\begin{equation}
\phi(x)=\phi_{c}+\varphi(x), \label{split}
\end{equation}
the next-to-last term in Eq.~(\ref{Lag}) gives a mass term for $A_{\mu}$:
 \begin{equation}
\Delta \mathcal{L}= -{1\over 2} f^2 (\alpha-1)\phi^2A_{\mu}^2 = {1\over 2} m_{A}^{2}A_{\mu}A^{\mu}+\cdot\cdot\cdot,
\end{equation}
where the mass squared parameter is given by
\begin{equation}
m_{A}^2= -f^2(\alpha-1)\phi_{c}^2 . \label{MA}
\end{equation}
Thus, for $m_{\phi}^2<0$, the Weyl gauge field $A_{\mu}$ acquires a positive mass squared only for timelike $\omega_{\mu}$. This can also be seen from the equation of motion of $A_{\mu}$, given by
\begin{equation}
\partial^{\mu} F_{\mu\nu}+f(\alpha -1) \left[{1\over 2} \partial_{\nu}\phi^2-f \phi^2 \omega_{\nu} - f \phi^2 A_{\nu}       \right]=0.
\end{equation}
 If $\langle\phi \rangle\neq0$, then the last term shows explicitly that $ m_{A}^2= -f^2(\alpha-1)\langle \phi \rangle^2  $ for $\alpha<1$, indicating that $\omega_{\mu}$ is timelike since $m_{\phi}^2<0$ in this case.

It is apparent that while $m_{\phi}^2$ is of order $f^2(\alpha-1)$, $m_{A}^2$ generated from the above mechanism is of order ${f^4\over\lambda}(\alpha-1)^2$. Also, it is conceivable that if the gauge fields of any compact Lie group are included and couple to the scalar field in the model, spontaneous gauge symmetry breaking can be triggered after Weyl symmetry breaking. Probing this scenario phenomenologically would be interesting but lies beyond the scope of this work.

What we have shown in this analysis is that when spontaneous breakdown of Weyl symmetry generates a mass for the scalar field, it does not guarantee the mechanism of mass generation for the Weyl gauge field. From now on, unless otherwise specified, we will focus on the case $m_{\phi}^2<0$ and assume $\omega_{\mu}$ to be timelike, so that $A_{\mu}$ obtains a mass from the breakdown of $Z_2$ symmetry.

\section{Naturalness}
One of the motivations for exploring Weyl invariant theories is to study how mass scales can be generated from a theory of no intrinsic scales and remain stable. This point was made in \cite{Oda2}. In the previous section, we have shown that for $m_{\phi}^2<0$ and timelike $\omega_{\mu}$, after Weyl symmetry breaking, another mass scale $m_A$ besides $m_\phi$ appears through the spontaneous breaking of the $Z_2$ symmetry for $\phi$. From the tree-level mass formula~(\ref{MA}), we have
\begin{equation}
m_{A}^2= {6f^2\over\lambda}(\alpha-1)m_{\phi}^2 . \label{MAA}
\end{equation}
For small values of $(\alpha -1)$, the relation~(\ref{MAA}) indicates that after symmetry breaking the theory defined by Eq.~(\ref{Action}) contains two hierarchically separated scales $m_{A} \ll m_{\phi}$ for couplings of $\lambda \sim O(1)$ and $f\sim O(1)$.

A theory possessing a hierarchy structure is commonly regarded as unnatural, and one might worry that the fine-tuning issue would appear since the interaction of the vector field $A_{\mu}$ with the scalar field $\phi$ is likely to introduce large quantum corrections to the mass of $A_{\mu}$. However, this is not the case.

To be explicit, let us assume that $\lambda \sim O(1)$ and $f\sim O(1)$ and analyze the quantum effect. From the interaction terms $-{1\over 2} f (\alpha-1)\phi^2(\partial_{\mu}+2f \omega_{\mu} )A^{\mu}$ and $ -{1\over 2} f^2 (\alpha-1)\phi^2 A_{\mu}^2$ in Eq.~(\ref{Lag}), it is straightforward to verify that due to the interaction of $A_{\mu}$ with $\phi$, the radiative correction from the $\phi$ loop to the mass squared for $A_{\mu}$ is
\begin{equation}
\delta m_{A}^2\sim {f^2\over 4\pi^2}(\alpha-1)m_{\phi}^2,
\end{equation}
which is a rather modest correction, compared with the tree-level result in Eq.~(\ref{MAA}).

The above result shows that the light scale $m_{A}$ is not sensitive to the heavy scale and the theory is technically natural. As we discussed in Sect. 2, this is due to the fact that the Lagrangian~(\ref{Lag}) possesses an enhanced Poincar$\rm\acute{e}$ symmetry in the limit $\alpha \rightarrow 1$. It is this extended spacetime symmetry that ensures the stability of the light scale by protecting it from large radiative corrections.

Surely in the above argument, we have neglected gravitational effects. However, it is credible that at energies well below the Planck scale, gravitational effects only give rise to small corrections and will not destabilize the hierarchy.

\section{Effective potential}
In Sect. 2, we have shown that for $\alpha <1$ and $\omega_{\mu}$ timelike, the breakdown of Weyl symmetry generates a negative mass squared for the scalar field $\phi$. Since the tree-level vacuum expectation value $\phi_c$ can be modified by perturbative loop corrections, in this section we would like to consider the lowest-order quantum corrections and determine the vacuum expectation value of the quantum field $\phi$.

We will follow the standard procedure by starting with the computation of the effective action, denoted by $\Gamma$. After obtaining $\Gamma$, we will identify an expression for the effective potential $V_{eff}$, whose minimum determines the quantum corrections to spontaneous symmetry breaking.

To calculate the effective potential to one-loop order, we follow the Faddeev-Popov method to gauge-fix the local symmetry and use Feynman gauge in our calculation. Including the gauge fixing term in Eq.~(\ref{Lag}), the gauge-fixed Lagrangian in the flat limit is of the form
\begin{equation}
{\cal{L}}_{FP}={\cal{L}}-{1\over2}(\partial_{\mu}A^{\mu})^2,
\end{equation}
Next, we split the scalar field $\phi (x)$ into a fixed classical part $\phi_c$ (we assume that translational symmetry remains intact) and a quantum fluctuation $\varphi (x)$, given by Eq.~(\ref{split}). For our purposes, we then expand the gauge-fixed Lagrangian ${\cal{L}}_{FP}$ about $\phi_c$ and keep pnly the terms up to quadratic order in quantum fields, We obtain
\begin{eqnarray}
{\cal{L}}_{2}&=&{1\over 2}\partial_{\mu} \varphi(x) \partial^{\mu} \varphi(x) -{1\over 2} m_{\phi}^{2} (\phi_{c}^2+\varphi^2(x)) - {1\over 4!}\lambda (\phi_{c}^4+6\phi_{c}^2\varphi^2(x)+\cdot\cdot\cdot ) \nonumber\\
& &-{1\over2}(\partial_{\mu}A^{\mu}(x))^2-{1\over 2} f^2 (\alpha-1)\varphi^2(x) A_{\mu}^2(x)-{1\over 4} F_{\mu\nu}^{2}(x). \label{L2}
\end{eqnarray}
Note that in ${\cal{L}}_2$ we have dropped terms linear in quantum fields by applying the classical field equations. From ${\cal{L}}_2$, we have
\begin{eqnarray}
-{\delta^2{\cal{L}}_{2}\over \delta A_{\mu}\delta A_{\nu}}&=&-\square \eta^{\mu\nu} + f^2 (\alpha-1)\phi_{c}^{2}  \eta^{\mu\nu},\\
-{\delta^2{\cal{L}}_{2}\over\delta \varphi \delta \varphi}& =& \square +m_{\phi}^2+{\lambda \over 2} \phi_{c}^{2}.
\end{eqnarray}
Then, using the background field method, it is straightforward to show that to one-loop order, the effective action $\Gamma[\phi_c]$ is given by
\begin{equation}
\Gamma[\phi_c]=\int d^4 x \left(   -{1\over 2}m_{\phi}^{2} \phi_{c}^2 -{1\over 4!}\lambda \phi_{c}^4  +\delta {\cal{L}}[\phi_{c}]\right)+2i\, {\rm log}\,{\rm{det}}\left[ \square- f^2 (\alpha-1)\phi_{c}^{2}\right]+{i\over 2} {\rm log}\,{\rm{det}}\left[ \square+m_{\phi}^2+{\lambda \over 2} \phi_{c}^{2}\right],
\end{equation}
where $\delta {\cal{L}}[\phi_{c}]$ contains all the counterterms to be determined by the renormalization conditions. Notice that for our purpose here we only include the $\phi_c$-dependent part of the effective action.

We remark that each functional determinant appears in $ \Gamma[\phi_c]$ is nothing but the determinant of a Klein-Gordon operator with $\phi_c$-dependent mass squared. This field dependence is due to the interaction of the propagating particles with the background field $\phi_c$. We also note that in the Faddeev-Popov procedure, it is important to include Faddeev-Popov ghosts which serve as negative degrees of freedom and render the counting of physical degrees of freedom correct. However, similar to QED, the Faddeev-Popov determinant is independent of $\phi_c$ in our case, and therefore it does not contribute to the effective potential $\Gamma[\phi_c]$.

To compute the functional determinant, we adopt dimensional regularization to evaluate the integral in $d=4-\epsilon$ dimensional spacetime. After this manipulation, we have
 \begin{eqnarray}
    {\rm log}\,{\rm{det}}\left[ \square+\mu^2(\phi_{c})\right]   &=&    {\rm Tr}\,{\rm log}\left[  \square+\mu^2(\phi_{c})\right]  =-i  (VT){ \Gamma({-{d\over2}})\over (4\pi)^{d/2}}( \mu^2(\phi_{c}) )^{d/2}         \nonumber\\
&=& -i(VT){\mu^4(\phi_c)\over 2(4\pi)^2}\left({2\over \epsilon}-\gamma+{\rm{log}}\,(4\pi)-{\rm{log}}\, (\mu^2(\phi_c)) +{3\over 2}  \right)    ,
\end{eqnarray}
where $VT$ is the spacetime volume of the functional integral.

After absorbing the divergences into the counterterms $\delta\cal{L}$ using the modified minimal subtraction scheme, we obtain the effective potential to one-loop order:
\begin{eqnarray}
V_{eff} =-{1\over VT}\Gamma[\phi_c]=-{1\over 2} \vert m_{\phi}^{2}\vert^2 +{\lambda\over 4!} \phi_{c}^{4}& + &{(-f^2(\alpha-1)\phi_{c}^{2})^2\over (4\pi)^2}\left( {\rm{log}}\, \left[ {-f^2(\alpha-1)\phi_{c}^{2}}\over M^2 \right]-{3\over 2}   \right)\nonumber\\
& + &{(m_{\phi}^2+{\lambda\over 2}\phi_{c}^2)^2\over4 (4\pi)^2}\left( {\rm{log}}\, \left[ m_{\phi}^2+{\lambda\over 2}\phi_{c}^2\over M^2 \right]-{3\over 2}   \right),
\end{eqnarray}
where $M$ is an arbitrary renormalization scale. It follows that the one-loop renormalization group $\beta$-function $\beta(\lambda)$ for $\lambda$ and the anomalous dimension $\gamma_{\phi_{c}^2} $ of the operator $\phi_{c}^2$ are, respectively,
\begin{eqnarray}
\beta(\lambda)&=&{3\over 16\pi^2}\left( \lambda^2+16 f^4 (\alpha-1)^2 \right),\\
\gamma_{\phi_{c}^2}&=&{\lambda \over 16\pi^2}.
\end{eqnarray}

For $V_{eff}$ to make any sense, the arguments of the logarithms must be positive. Thus, for $\alpha<1$ and $\omega_{\mu}$ timelike, the quantum correction to $V_{eff}$ is consistent only if
\begin{equation}
m_{\phi}^2+{\lambda\over 2}\phi_{c}^2 >0, \label{mass}
\end{equation}
that is,
\begin{equation}
\phi_{c}^2 > {2 f^2(1-\alpha) \over \lambda} \omega_{\mu}^2 .
\end{equation}

Expanding $V_{eff}$ in powers of $\vert m_{\phi}^{2}\vert$, we have
\begin{eqnarray}
V_{eff}&=& {\lambda\over 4!} \phi_{c}^{4} + {(f^2(\alpha-1)\phi_{c}^{2})^2\over (4\pi)^2}\left( {\rm{log}}\, \left[ {-f^2(\alpha-1)\phi_{c}^{2}}\over M^2 \right]-{3\over 2}   \right)+ {(\lambda\phi_{c}^2)^2\over 16 (4\pi)^2}\left( {\rm{log}}\, \left[{\lambda\phi_{c}^2\over 2 M^2} \right]-{3\over 2} \right) \nonumber\\
&- & {1\over 2} \vert m_{\phi}^{2}\vert \phi_{c}^2 \left(  1+ {\lambda\over 2 (4\pi)^2}\left( {\rm{log}}\,\left[ {\lambda\over 2}{\phi_{c}^{2}\over M^2}  \right]-1   \right)  \right)+{\rm{O}}(m_{\phi}^4).
\end{eqnarray}
The above result shows that for small values of $\vert m_{\phi}^{2}\vert$, the one-loop contributions to the effective potential takes the form
\begin{equation}
V_{eff}^{(1)}=a_{1} \phi_{c}^2 +a_{2} \phi_{c}^4+{1\over 16\pi^2} \left[ \left(f^4(\alpha-1)^{2}+{\lambda^2\over 16}\right) \phi_{c}^4\, {\rm{log}}\, \phi_{c}^{2}   -{\lambda\over 4} \vert m_{\phi}^{2}\vert \phi_{c}^{2}  \,  {\rm{log}}\, \phi_{c}^{2} \right] + {\rm{O}}(m_{\phi}^4).
\end{equation}
Imposing the following renormalization conditions on the one-loop effective potential:
\begin{equation}
V_{eff}^{(1)\,''''}\vert_{\phi_{c}=\mu}=0,\,\,\,{\rm{and}}\,\,\,\,V_{eff}^{(1)\,''}\vert_{\phi_{c}=\mu}=0, \label{ren}
\end{equation}
we find
\begin{eqnarray}
a_{1}&=& {1\over 16\pi^2}\left( 18 \left(f^4(\alpha-1)^2+{\lambda^2\over 16}\right)\mu^2 +\lambda\vert m_{\phi}^{2}\vert +{\lambda\over 4}\vert m_{\phi}^{2}\vert \, {\rm{log}}\, \mu^2 \right)   ,\\
a_{2}&=& -{1\over 16\pi^2}\left(  \left(f^4(\alpha-1)^2+{\lambda^2\over 16}\right)\left( {\rm{log}}\, \mu^2+{25\over 6}    \right)   +{\lambda\over 24}{\vert  m_{\phi}^{2}\vert \over \mu^2}    \right).
\end{eqnarray}
Next, following Coleman and Weinberg \cite{Coleman} to choose $\mu=\langle \phi \rangle$ with $\langle\phi\rangle$ at the local minimum of the effective potential\footnote{With this choice, the renormalization condition for $V_{eff}^{(1)\,''}$ is equivalent to $V_{eff}^{'' }\vert_{\phi_{c}=\langle \phi\rangle}=(V_{eff}^{(0)\,''}+ V_{eff}^{(1)\,''})\vert_{\phi_{c}=\langle \phi\rangle}= V_{eff}^{(0)\,''}\vert_{\phi_{c}=\langle \phi\rangle}=-\vert m_{\phi}^2\vert +{\lambda\over 2} \langle \phi \rangle^2$. This guarantees that $\langle \phi \rangle$ is a local minimum, because of Eq.~(\ref{mass}).}, and then including the classical potential, we finally have
\begin{eqnarray}
V_{eff}(\phi_{c})&=&{1\over 2} \left[ {9\over 4\pi^2}\left(f^4(\alpha -1)^2+{\lambda^2\over 16}\right)\langle \phi \rangle^2  -\left( 1+{\lambda \over 8\pi^2}   \right) \vert m_{\phi}^2\vert \right] \phi_{c}^{2}+   {\lambda\over 4!}\left(1-{1\over 16\pi^2}{ \vert m_{\phi}^2 \vert \over \langle \phi \rangle^2}   \right)\phi_{c}^4 \nonumber \\
&+&{1\over 16\pi^2}\left[\left(f^4(\alpha -1)^2+{\lambda^2\over 16}\right)\phi_{c}^{4}\left(  {\rm{log}}\left({\phi_{c}^2\over \langle \phi \rangle^2}\right)-{25\over 6}   \right)-{\lambda\over 4}\vert m_{\phi}^2\vert \phi_{c}^{2}\,{\rm{log}}\left({\phi_{c}^2\over \langle \phi \rangle^2}\right)   \right]\nonumber\\
&+&{\rm{O}}(m_{\phi}^4). \label{V}
\end{eqnarray}
It is then straightforward to show that $V_{eff}$ has its minimum at
\begin{equation}
\langle \phi \rangle^2 ={(1+{\lambda\over 6\pi^2})\over {\lambda\over 6}+{4\over 3\pi^2}\left(f^4(\alpha -1)^2+{\lambda^2\over 16}\right)}\vert m_{\phi}^{2}\vert, \label{min}
\end{equation}
which is away from the origin as long as $\vert m_{\phi}^{2}\vert \neq 0$. This is to be compared to the tree-level result $\langle \phi \rangle_{tree}^2=6\vert m_{\phi}\vert^{2}/\lambda$.

In the final part of this section, let us consider the case where $m_{\phi}^{2}=0$ (i.e., $\omega_{\mu}$ lightlike) in Eq.~(\ref{Lag}). Naively, one might think that the effective potential~(\ref{V}) and the location of its minimum~(\ref{min}) could be applied to the case where $\omega_{\mu}$ is lightlike by simply taking $m_{\phi}=0$ in Eq.~(\ref{V}) and Eq.~(\ref{min}). However, this is incorrect. The reason is that when $m_{\phi}=0$, the renormalization condition for $V_{eff}^{(1)\,''}$ in Eq.~(\ref{ren}) implies that
\begin{equation}
V_{eff}^{''}\vert_{\phi_{c}=\langle\phi\rangle}={\lambda\over 2} \langle \phi\rangle^2.
\end{equation}
This result is consistent with the assumption that $\langle\phi\rangle$ is at a local minimum of $V_{eff}$ only if $\langle\phi\rangle \neq 0$ (so that $V_{eff}^{''}\vert_{\phi_{c}=\langle\phi\rangle}>0$). Nevertheless, Eq.~(\ref{min}) tells us that when $m_{\phi}=0$, $\langle\phi\rangle$ is away from the origin only if ${\lambda\over 6}+{4\over 3\pi^2}\left(f^4(\alpha -1)^2+{\lambda^2\over 16}\right)=0$, which cannot be satisfied for $\lambda>0$. Therefore, we need to adopt another renormalizaton condition for $ V_{eff}^{(1)\,''}$ when $m_{\phi}=0$.

To do this, let $V_{0,eff}$ denote the effective potential for the $m_{\phi}=0$ case, and let $\langle\phi\rangle $ again denote the location of the minimum of $V_{0,eff}$. Then, following the same manipulations as in the $\vert m_{\phi}^{2}\vert\neq 0$ case, except choosing new renormalization conditions:
\begin{equation}
V_{0,eff}^{(1)\,''''}\vert_{\phi_{c}=\langle\phi\rangle}=0,\,\,\,{\rm{and}}\,\,\,\,V_{0,eff}^{(1)\,''}\vert_{\phi_{c}=0}=0,
\end{equation}
we find
\begin{equation}
V_{0,eff}(\phi_{c})={\lambda\over 4!}\phi_{c}^{4}+{1\over 256\pi^2}\left(16f^4(\alpha -1)^2+\lambda^2\right)\phi_{c}^{4} \left({\rm{log}}\left({ \phi_{c}^2 \over \langle\phi\rangle^2}  \right)-{25\over 6}  \right). \label{V0}
\end{equation}
Notice that when $\alpha=1$, The above expression reduces to the effective potential of the conformal scalar theory to one-loop order. This should be an expected result, after the discussion leading to Eq.~(\ref{SF}) in Sect. 2.

From Eq.(\ref{V0}), it is easy to show that $\langle\phi\rangle $ is away from the origin only if
\begin{equation}
f^4(\alpha -1)^2+{\lambda^2\over 16}={2\pi^2\over 11}\lambda, \label{lambda}
\end{equation}
which is consistent with the validity of the perturbation theory if $\lambda$ is of order $f^4 (\alpha-1)^2$. With the value of $\lambda$ satisfying Eq.~(\ref{lambda}), the effective potential $V_{0,eff}$ can be re-expressed as
\begin{equation}
V_{0,eff}(\phi_{c})={\lambda\over 88}\phi_{c}^{4} \left({\rm{log}}\left({ \phi_{c}^2 \over \langle\phi\rangle^2}  \right)-{1\over 2}  \right). \label{V00}
\end{equation}
The above result exhibits dimensional transmutation such that the dimensionless quantity $f^2(\alpha-1)$ is traded for the dimensional scale $\langle\phi\rangle$.

The masses of $\phi$ and $A_{\mu}$ are thus given by
\begin{equation}
m_{\phi}^2=V_{0,eff}^{''}\vert_{\langle\phi\rangle}={\lambda\over 11}\langle\phi\rangle^2\simeq{f^4(\alpha-1)^2\over 2\pi^2}\langle\phi\rangle^2,
\end{equation}
and
\begin{equation}
m_{A}^2=f^2(1-\alpha)\langle\phi\rangle^2 ={11f^2(1-\alpha)\over \lambda} m_{\phi}^2,
\end{equation}
which is physically sensible for $\alpha <1$.

This analysis justifies that for lightlike $\omega_{\mu}$, the $Z_{2}$ symmetry of the $\phi$ field can be broken dynamically via Coleman-Weinberg mechanism and from which the masses of fields are generated.

\section{Conclusion}
Probing the possible violation of Lorentz symmetry is an interesting subject both theoretically and experimentally. In this paper, we investigated the scenario in which Lorentz violation is realized by a non-vanishing vacuum expectation value of the Weyl field after Weyl symmetry breaking. In particular, a specific scalar field theory with Weyl invariance has been studied. Characteristic features of different symmetry-broken phases after Weyl symmetry breaking and their dependence on the spacetime character of the Lorentz-violating background were discussed. We found that in either a lightlike or a timelike Lorentz-violating background, the Weyl field could acquire a positive mass-squared after Weyl symmetry breaking. In the lightlike case, this is achieved via dynamical symmetry breaking. From the point of view of effective field theories, the criticisms on Weyl geometry no longer hold at low energies after integrating out the massive Weyl field.

A Weyl invariant theory also serves as an intriguing model for one to explore and gain insights into not only the origin but also the stability of separated mass scales. We have analyzed the naturalness of the theory. The upshot is that, due to the enhanced Poincar$\rm{\acute{e}}$ symmetry in the limit $\alpha\rightarrow 0$, the light mass-squared parameter receives no large quantum corrections.

Based on these results, keeping in mind that the Weyl field does not couple to the gauge and fermion fields in the SM, further explorations of this scenario and its possible phenomenological applications in physics beyond the SM and in cosmology are appealing. For example, following the approach of this work, one may consider the extension of the conformal SM in Weyl geometry such that the dimensionful parameter of the Higgs sector is generated after Weyl symmetry breaking, similar to the model we studied in this work. Since only the scalar sector couples to the Weyl gauge field, the data of precision Higgs boson measurements combined with the current lower bound on the non-metricity scale would impose constraints on the dimensionless parameters of the model. We hope to study this in the near future.

\section*{Acknowledgments}
\hspace*{\parindent}

This research was supported in part by the National Nature Science Foundation of China under Grant No. 11665016.



\end{document}